\documentclass[aps,prc,superscriptaddress,showpacs,amsmath,amssymb,twocolumn]{revtex4-1}
\usepackage{graphicx}% Include figure files
\usepackage{dcolumn}% Align table columns on decimal point
\usepackage{multirow}
\usepackage{bm}% bold math
\usepackage{tensor}
\usepackage{textcomp}
\usepackage{color}
\usepackage{threeparttable}
\usepackage[colorlinks,allcolors=black]{hyperref}  
\newcommand{\Cag}{\ensuremath{{}^{12}\mathrm{C}(\alpha,\gamma){}^{16}\mathrm{O}}}
\newcommand{\Can}{\ensuremath{{}^{13}\mathrm{C}(\alpha,\mathrm{n}){}^{16}\mathrm{O}}}
\newcommand{\Clid}[1]{\ensuremath{{}^{12}\mathrm{C}({}^{6}\mathrm{Li},\mathrm{d}){}^{16}\mathrm{O}_{#1}}}
\newcommand{\LiCd}{\ensuremath{{}^{6}\mathrm{Li}({}^{12}\mathrm{C},\mathrm{d}){}^{16}\mathrm{O}}}
\newcommand{\Clit}[1]{\ensuremath{{}^{12}\mathrm{C}({}^{7}\mathrm{Li},\mathrm{t}){}^{16}\mathrm{O}_{#1}}}
\newcommand{\Cbli}{\ensuremath{{}^{12}\mathrm{C}({}^{11}\mathrm{B},{}^{7}\mathrm{Li}){}^{16}\mathrm{O}}}
\newcommand{\lilidgs}{\ensuremath{{}^{7}\mathrm{Li}({}^{6}\mathrm{Li},\mathrm{d}){}^{11}\mathrm{B}_\mathrm{g.s.}}}

\begin{document}
\preprint{APS/123-QED}

\title{Astrophysical $S_{E2}$ factor of the \Cag{} reaction through the \Cbli{} transfer reaction}



\author{Y.~P.~Shen}
\affiliation{China Institute of Atomic Energy, P. O. Box 275(10), Beijing 102413, China}
\author{B.~Guo}
\email[]{guobing@ciae.ac.cn}
\affiliation{China Institute of Atomic Energy, P. O. Box 275(10), Beijing 102413, China}
\author{Z.~H.~Li}
\affiliation{China Institute of Atomic Energy, P. O. Box 275(10), Beijing 102413, China}
\author{Y.~J.~Li}
\affiliation{China Institute of Atomic Energy, P. O. Box 275(10), Beijing 102413, China}
\author{D.~Y.~Pang}
\affiliation{School of Physics and Nuclear Energy Engineering, Beihang University, Beijing 100191, China}
\affiliation{Beijing Key Laboratory of Advanced Nuclear Materials and Physics, Beihang University, Beijing 100191, China}
\author{S.~Adhikari}
\affiliation{Physics Department, Techno India University, Kolkata 700091, India}
\author{Z.~D.~An}
\affiliation{School of Physics and Astronomy, Sun Yat-sen University, Zhuhai 519082, China}
\author{J.~Su}
\affiliation{China Institute of Atomic Energy, P. O. Box 275(10), Beijing 102413, China}
\author{S.~Q.~Yan}
\affiliation{China Institute of Atomic Energy, P. O. Box 275(10), Beijing 102413, China}
\author{X.~C.~Du}
\affiliation{China Institute of Atomic Energy, P. O. Box 275(10), Beijing 102413, China}
\author{Q.~W.~Fan}
\affiliation{China Institute of Atomic Energy, P. O. Box 275(10), Beijing 102413, China}
\author{L.~Gan}
\affiliation{China Institute of Atomic Energy, P. O. Box 275(10), Beijing 102413, China}
\author{Z.~Y.~Han}
\affiliation{China Institute of Atomic Energy, P. O. Box 275(10), Beijing 102413, China}
\author{D.~H.~Li}
\affiliation{China Institute of Atomic Energy, P. O. Box 275(10), Beijing 102413, China}
\author{E.~T.~Li}
\affiliation{College of Physics and Energy, Shenzhen University, Shenzhen 518060, China}
\author{X.~Y.~Li}
\affiliation{China Institute of Atomic Energy, P. O. Box 275(10), Beijing 102413, China}
\author{G.~Lian}
\affiliation{China Institute of Atomic Energy, P. O. Box 275(10), Beijing 102413, China}
\author{J.~C.~Liu}
\affiliation{China Institute of Atomic Energy, P. O. Box 275(10), Beijing 102413, China}
\author{T.~L.~Ma}
\affiliation{China Institute of Atomic Energy, P. O. Box 275(10), Beijing 102413, China}
\author{C.~J.~Pei}
\affiliation{China Institute of Atomic Energy, P. O. Box 275(10), Beijing 102413, China}
\author{Y.~Su}
\affiliation{China Institute of Atomic Energy, P. O. Box 275(10), Beijing 102413, China}
\author{Y.~B.~Wang}
\affiliation{China Institute of Atomic Energy, P. O. Box 275(10), Beijing 102413, China}
\author{S.~Zeng}
\affiliation{China Institute of Atomic Energy, P. O. Box 275(10), Beijing 102413, China}
\author{Y.~Zhou}
\affiliation{China Institute of Atomic Energy, P. O. Box 275(10), Beijing 102413, China}
\author{W.~P.~Liu}
\affiliation{China Institute of Atomic Energy, P. O. Box 275(10), Beijing 102413, China}


\date{\today}

\begin{abstract}
The \Cag{} reaction plays a key role in the evolution of stars with masses of $M >$ 0.55 $M_\odot$. The cross-section of the \Cag{} reaction within the Gamow window ($E_\textrm{c.m.}$ = 300 keV, $T_\textrm9$ = 0.2) is extremely small (about $10^{-17}$ barn), which makes the direct measurement in a ground-based laboratory with existing techniques unfeasible. Up until now, the cross-sections at lower energies can only be extrapolated from the data at higher energies. However, two subthreshold resonances, located at $E_x$ = 7.117 MeV and $E_x$ = 6.917 MeV, make this extrapolation more complicated. In this work, the 6.917 MeV subthreshold resonance in the \Cag{} reaction was investigated via the \Cbli{} reaction. The experiment was performed using the Q3D magnetic spectrograph at the HI-13 tandem accelerator. We measured the angular distribution of the \Cbli{} transfer reaction leading to the 6.917 MeV state. Based on the FRDWBA analysis, we derived the asymptotic normalization coefficient (ANC) of the 6.917 MeV level in $^{16}$O to be (1.10 $\pm$ 0.29) $\times 10^{10}$ fm$^{-1}$, with which the reduced $\alpha$ width was computed to be $18.0\pm4.7$ keV at the channel radius of 6.5 fm. Finally, we calculated the astrophysical $S_{E2}(300)$ factor of the ground-state transitions to be 46.2 $\pm$ 7.7 keV b. The result for the astrophysical $S_{E2}(300)$ factor confirms the values obtained in various direct and indirect measurements and presents an independent examination of the most important data in nuclear astrophysics.
\end{abstract}


\maketitle

\section{Introduction}
\label{intro}

The \Cag{} reaction is believed to be one of the most crucial reactions in nuclear astrophysics \cite{wea93,wal97,sch12}. Following the production of $^{12}$C by the triple-$\alpha$ process, it strongly influences the ratio of the abundances for the main isotopes of carbon and oxygen ($^{12}$C and $^{16}$O) which are the fourth- and third-most abundant nuclei in the visible universe. The C/O ratio at the end of helium burning affects not only the production of all elements heavier than $A = 16$, but also the explosion of supernovae \cite{wea93}. While the cross-section for the triple-$\alpha$ process is experimentally well determined \cite{han05,buc06}, the cross-section of the \Cag{} reaction taking place in the helium burning phase ($T_9$ = 0.2) is now thought to be with the most serious uncertainty in nucleosynthesis \cite{fow84}. The center of the Gamow peak for the \Cag{} reaction at $T_9$ = 0.2 is located at $E_\textrm{c.m.}$ = 300 keV. Stellar modeling requires the uncertainty for the \Cag{} cross-section at $E_\textrm{c.m.}$ = 300 keV to be better than 10\% \cite{wea93,woo02}, while the present uncertainty is approximately 20\% \cite{xu13}.

At energies corresponding to the Gamow peak, the $^{12}$C($\alpha$,\,$\gamma$)$^{16}$O cross-sections are too low (on the order of 10$^{-17}$ barn) to be measured directly in a ground-based laboratory. Although, in the near future, a direct measurement is planned by the JUNA collaboration \cite{liu16}, all direct measurements thus far have been done at energies higher than $E_{\textrm{c.m.}}$ = 890 keV \cite{fey04,ham05,ham05b}. How to achieve a reliable extrapolation of the cross-sections from present data to the Gamow window has been a long-standing problem. Furthermore two subthreshold resonances, 7.117 MeV 1$^-$ and 6.917 MeV 2$^+$, make this extrapolation more complicated. There are two main capture modes in the \Cag{} reaction. One is the $E1$ transition to the ground state that includes the contributions from the low-energy tail of the broad 1$^-$ resonance at $E_x$ = 9.585 MeV and the subthreshold 1$^-$ resonance at $E_x$ = 7.117 MeV. The other is the $E2$ transition to the ground state, which mainly stems from the direct capture and the subthreshold 2$^+$ resonance at $E_x$ = 6.917 MeV. The states with identical multipolarity interfere with each other. $R$-matrix analysis is a widely used method to deal with situations which require level parameters (i.e., energies, ANCs and lifetimes). Indirect techniques are believed to be quite valuable since they can be used to deduce these level parameters \cite{deb17}. To date, considerable indirect methods have been utilized to study these two subthreshold resonances, such as the $\alpha$ + $^{12}$C elastic scattering \cite{tis09}, the $\beta$-delayed $\alpha$ decay of $^{16}$N \cite{ref16}, transfer reactions \cite{adh14} and Coulomb dissociation \cite{fle05}. All of these results for the $S_{E2}(300)$ factor vary from 36 to 85 keV b.

Due to the importance of the transfer method to evaluate the \Cag{} reaction, a lot of work has been done \cite{puh70,joh70,cob76,bec78,bec78b,cun78,bec80,bru99,dru99,kee03,bel07,adh11,oul12,adh14,avi15} to date. 
P\"uhlhofer et al. \cite{puh70} measured the angular distributions of \Clit{} at $E_{lab}=15, 21.1$ and $24$ MeV and the reduced $\alpha$-widths of some states were extracted. Johnson et al. \cite{joh70} measured the reaction \Clid{} within an energy range from 5.6 to 14.0 MeV. However, the events of \Clid{6.92} could not be distinguished from the events of \Clid{7.12} and no further analysis such as DWBA was made. The \Clit{} was measured in Cobern et al.'s \cite{cob76} work, but they were also unable to tell the events from 6.92 MeV and 7.12 MeV apart. Cunsolo et al. measured the \Clid{} reaction in the 20-34 MeV incident energy range and analyzed in terms of Hauser-Feshbach and FRDWBA theories. Becchetti et al. \cite{bec78,bec78b} measured the \Clid{} and \Clit{} in 1978 at energies of 42 MeV and 34 MeV, respectively. In 1980, Becchetti et al. \cite{bec80} measured \Clid{} again at 90 MeV. The Hauser-Feshbach and FRDW theories were applied in the analysis. Brune et al. \cite{bru99} measured the \Clid{} and \Clit{} reactions to the bound $2^+$ and $1^-$ states of $^{16}$O and analyzed these data using the finite-range DWBA code FRESCO \cite{tho88}. Drummer et al. \cite{dru99} measured the \Clid{g.s.} with a polarized $^6$Li beam. Keeley et al. \cite{kee03} measured \Clid{} at 34 and 50 MeV and analyzed the multistep contributions to the transfers leading to the $0^+$, $2^+$, $4^+$ and $3^-$ states. Belhout et al. \cite{bel07} measured \Clid{} and 48.2 MeV and analyzed it using the FRDWBA theory with a particular emphasis put on the states of astrophysical interest, mainly, the 7.12 MeV state. Oulebsir et al. \cite{oul12} measured \Clit{} reaction at two incident energies 28 and 34 MeV and analyzed this using the Hauser-Feshbash and FRDWBA theories. Adhikari et al. \cite{adh11,adh14} measured \Clid{} at 9 and 20 MeV, and continuum discretized coupled channel-coupled reaction channel (CDCC-CRC) calculations have been used to analyze the data. Avila et al. \cite{avi15} applied the $\alpha$-transfer reaction \LiCd{} with inverse kinematics and constrained the 6.05 MeV and 6.13 MeV cascade transitions in the \Cag{} reaction.
Particularly, the $S_{E2}(300)$ factors are extracted in \cite{bru99,bel07,oul12,adh14}. As mentioned above, all of the works are performed with the \Clid{} and \Clit{} transfer systems. It is known that one of the largest sources of uncertainty in the ANC determination from these studies is the uncertainty in the FRDWBA model. For this reason, measurements of different types of transfer reactions may help us to better understand the systematic uncertainties in the model and lead to improvement in the method. Additional measurement via independent transfer reactions is therefore desirable. In addition to the ($^6$Li,d) and ($^7$Li,t) reactions, the ($^{11}$B, $^7$Li) transfer reaction is another choice for research in ($\alpha$,$\gamma$) or ($\alpha$,n) reactions, which has been successfully applied to the research of \Can{} \cite{guo12}. In our present work, measurement of the \Cbli{} reaction was performed to derive the reduced $\alpha$ width of the 6.917 MeV 2$^+$ subthreshold resonance. The astrophysical $S_{E2}$ factor at the Gamow peak of the \Cag{} reaction was then studied.

\section{Experiment} % (fold)
\label{sec:experiment}

\begin{figure}[htbp]
\centering
\includegraphics[width=\columnwidth]{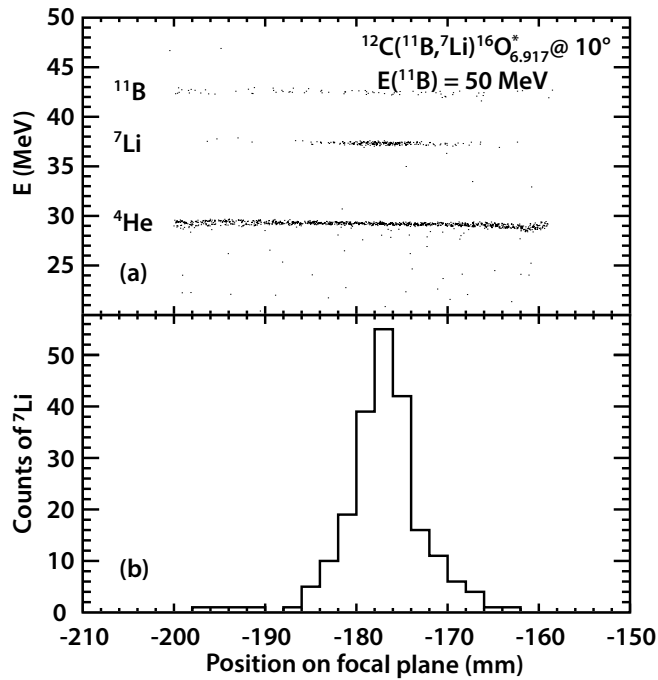}
\caption{Focal-plane position spectrum of $^7$Li at $\theta_\mathrm{lab}$\,=\,10$^\circ$ from the \Cbli{} reaction. (a) Two-dimensional spectrum of energy vs. focal-plane position. (b) Spectrum gated by the $^7$Li events in (a).}
\label{fig1}       % Give a unique label
\end{figure}

The experiment was performed at the HI-13 national tandem accelerator laboratory of the China Institute of Atomic Energy (CIAE) in Beijing. The experimental setup and procedures were similar to those previously reported \cite{guo12,guo14,li12,li13}. The $^{11}$B beam with an energy of 50 MeV was delivered and utilized to measure the angular distribution of the \Cbli{} reaction leading to the excited state of $^{16}$O at $E_x$ = 6.917 MeV and $^{11}$B+$^{12}$C elastic scattering. A self-supporting $^{12}$C target with a thickness of $66\pm5~\mathrm{\mu g/cm^2}$ was used in the present experiment. In addition, the $^7$Li beam with an energy of 26 MeV and a SiO$_2$ target with a thickness of $86\pm7~\mathrm{\mu g/cm^2}$ were used for the measurement of the angular distribution of the $^7$Li+$^{16}$O elastic scattering. The reaction products were focused and separated by the Q3D magnetic spectrograph. A two-dimensional position sensitive silicon detector (X1) was fixed at the focal plane of Q3D. The two-dimensional position information from X1 enabled the products emitted into the acceptable solid angle to be completely recorded, and the energy information was used to remove the impurities with the same magnetic rigidity. As an example, Fig. \ref{fig1} displays the focal-plane position spectrum of $^7$Li at $\theta_\mathrm{lab}$\,=\,10$^\circ$ from the \Cbli{} reaction. We found that the energy resolution is approximately 40 keV and the $^7$Li events related to the 6.917 MeV state are well separated from others. The events related to the 7.12 MeV state are about 40 mm away from the events related to the 6.917 MeV state. Since we used one piece of silicon detector with a length of 50 mm, it's difficult to measure the two states in one run. Thus, the data for the 7.117 MeV state are not presented in this work.

We found that the $^{12}$C will build up on the front surface of the target because of the oil vapor in the beam pipe \cite{guo16}. Apparently, the buildup of $^{12}$C will increase the amount of the $^{12}$C atoms in the target, and influence the determination of the reaction cross-sections. To monitor the possible buildup of $^{12}$C, the $^{11}$B elastic scattering on the $^{12}$C target was measured at the start and the end of the measurement for each angle. Although the amount of $^{12}$C in the target increased by about 9\% during the whole measurement, it was less than 2\% for the measurement of the cross-sections at a single angle. All the measured cross-sections were corrected for the change in target thickness and the uncertainty of target thickness from the buildup of $^{12}$C was also included in the present work to avoid unexpected systematic error.

\begin{figure}[htbp]
\centering
\includegraphics[width=\columnwidth]{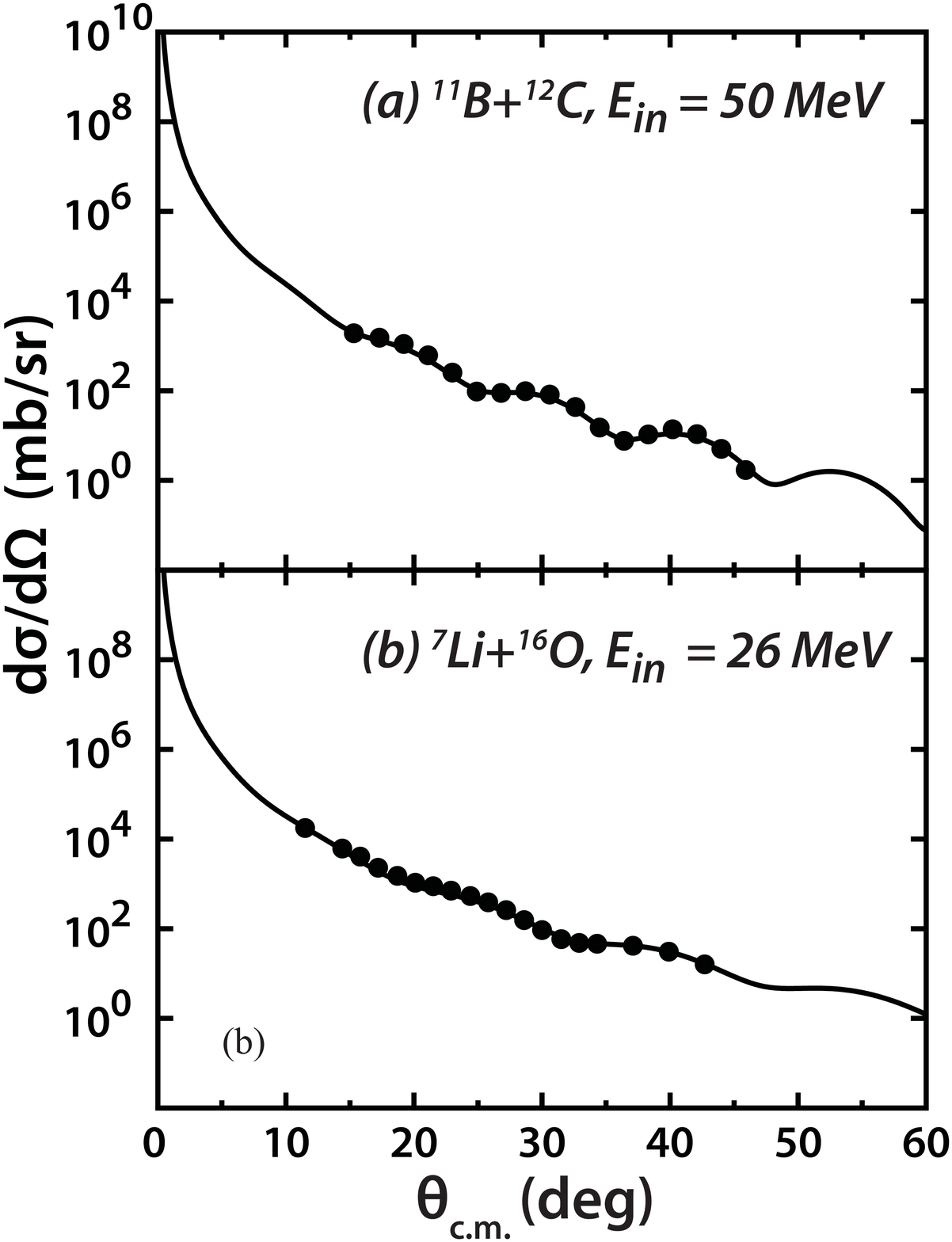}
\caption{The experimental data and fitting with single-folding potentials of the elastic scattering of $^{11}$B+$^{12}$C (showed in subplot (a)) and $^{7}$Li+$^{16}$O (showed in subplot (b)) which are the entrance and exit channels of \Cbli.}
\label{elscatt}       % Give a unique label
\end{figure}

\begin{figure}[htbp]
\centering
\includegraphics[width=\columnwidth]{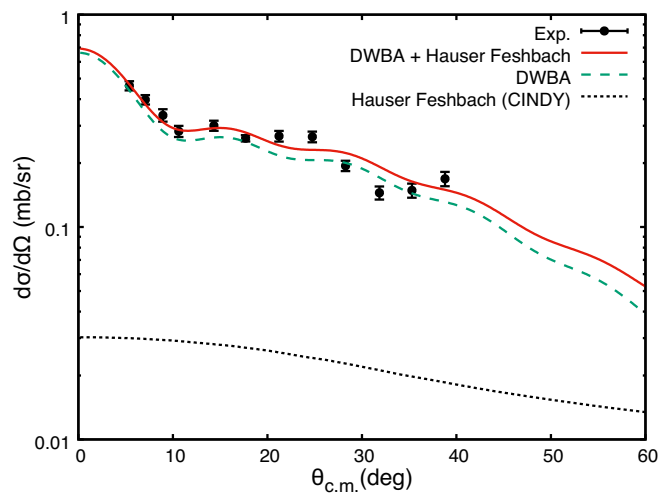}
\caption{Angular distribution of the \Cbli{} reaction leading to the 6.917 MeV 2$^+$ state of $^{16}$O. The green dashed line and the black dotted line represent the FRDWBA and Hauser-Feshbach calculation, respectively. The red solid line denotes the FRDWBA calculation summed to the compound nuclear component.}
\label{fig3}       % Give a unique label
\end{figure}

In order to derive the optical potential of the entrance and exit channels of the \Cbli{} reaction, we performed measurements of the elastic scattering of the $^{11}$B+$^{12}$C and $^{7}$Li+$^{16}$O at energies of 50 MeV and 26 MeV, respectively. The data for the differential cross-sections and fitting curves are shown in Fig. \ref{elscatt}. In Fig. \ref{fig3}, we also display the angular distribution of the \Cbli{} reaction leading to the 6.917 MeV 2$^+$ state of $^{11}$O. 

% section experiment (end)

\section{Extraction of ANC} % (fold)
\label{sec:extraction_of_anc}

The finite-range distorted wave Born approximation (FRDWBA) calculations were performed to derive the ANC of the 6.917 MeV 2$^+$ subthreshold state in $^{16}$O by using the FRESCO code \cite{tho88}. The FRDWBA calculations required the optical potentials for the entrance channel ($^{11}$B+$^{12}$C), exit channel ($^{7}$Li+$^{16}$O), and the core-core ($^{7}$Li+$^{12}$C) interactions. The real binding potentials for the ($\alpha$+$^{7}$Li) and ($\alpha$+$^{12}$C) systems were also required. As the depths of the binding potentials would be automatically adjusted during the calculation by the FRESCO code, only the geometrical parameters for the binding potentials were required.

A single-folding model \cite{pan11,xuy13} was used for the optical model potentials of the entrance and exit channels. Nucleon density distributions of $^{11}$B, $^{12}$C, and $^{16}$O were obtained using Hartree-Fock calculations with the SkX interaction \cite{bro98}, while those of $^7$Li were taken from a independent-particle model \cite{sat79}. These density distributions were folded using the systematic nucleon-nucleus potential of the JLMB model \cite{bau01}. Depths of these single-folding potentials were adjusted by normalizing parameters to provide an optimum reproduction of the experimental data with the optical model. The comparisons with the experimental data and the optical model calculations with these potentials are depicted in Fig. \ref{elscatt}. An approximation that we implemented was that the same optical potential was used for both the exit channel and the core-core interaction. The way to prove the correction of this approximation is to compare the difference between the prior and post interaction in the FRDWBA calculation. The difference in the present work was less than 1\%, which verified our approximation for the core-core optical potential. The uncertainties of the normalizing parameters of the single-folding potentials for entrance and exit channels were evaluated based on a least square minimization procedure and the total impact of the normalizing parameters on the $S_\alpha$ and the ANC was found to be approximately 6\%. The details of the normalization parameters and the uncertainties are shown in Table \ref{tab:uncertainties}.

The geomaetric parameters of binding potentials for the ($\alpha$+$^{7}$Li) and ($\alpha$+$^{12}$C) systems were another important input for the FRDWBA calculation. The geometric parameters, radius $r_0$ and diffuseness $a$ for the ($\alpha$+$^{7}$Li) system were adjusted to reproduce the root-mean-square (rms) radius ($\sqrt{\langle r^2\rangle}=3.204$ fm) of the $\alpha$-cluster wave function using the formula 
\begin{equation}
    \left\langle r_\mathrm{B}^2\right\rangle=\frac{m_{\mathrm{He}}}{m_\mathrm{B}}\left\langle r_{\mathrm{He}}^2\right\rangle+\frac{m_\mathrm{Li}}{m_\mathrm{B}}\left\langle r_\mathrm{Li}^2\right\rangle+\frac{m_\mathrm{He}m_\mathrm{Li}}{m_\mathrm{B}^2}\left\langle r^2\right\rangle
\end{equation}
given in Guo et al. (2012) \cite{guo12} where the rms radii of $^4$He, $^7$Li, and $^{11}$B were taken to be 1.47 fm, 2.384 fm, and 2.605 fm, respectively \cite{lia90}. The resulting parameters were $r_0=0.98$ fm and $a=0.60$ fm. We investigated the dependence of the calculated $S_\alpha$ and ANC on the geometric parameters for the ($\alpha$+$^{7}$Li) system. With a radius between 0.86-1.10 fm, the diffuseness was adjusted to reproduce the rms radius of 3.204 fm. The impact of this change on the ANC was found to be approximately 3\%. The geometric parameters for ($\alpha$+$^{12}$C) system were deduced following a similar procedure. While the rms radii of $^{4}$He, $^{12}$C and the first $2^+$ states of $^{16}$O are recommended to be 1.47 fm \cite{lia90}, 2.481 fm \cite{lia90} and 3.1 fm \cite{fun15}, respectively, the rms radius of the $\alpha$-cluster wave function was found to be 4.87 fm. The geometry parameters were deduced to be $r_0=0.96$ fm and $a=0.9$ fm. We varied the radius between 0.84-1.08 fm and adjusted diffuseness to reproduce the rms radius of 4.87 fm. The impact of the change on the ANC was found to be approximately 11\%. In the works of Oulebsir et al. (2012) \cite{oul12} and Keeley et al. (2003) \cite{kee03}, the geometric parameters of the ($\alpha$+$^{12}$C) system were recommended to be $r_0=1.16$ fm, $a=0.73$ fm and $r_0=1.25$ fm, $a=0.65$ fm, respectively. The above two sets of parameters only caused a change of less than 5\% on the ANC, which provided a crosscheck to our geometric parameters.

To obtain the spectroscopic factor and ANC of the $\alpha$-cluster in the $^{16}$O$_{6.917}$, the spectroscopic amplitudes of the $\alpha$-cluster in the ground state of $^{11}$B needed to be fixed. The single-particle wave function describing the relative motion between the $\alpha$-cluster and the $^7$Li core in the $^{11}$B ground state has two components denoted by quantum numbers $NL_j$ = 3$S_0$ and 2$D_2$, respectively. Another experiment was performed on our facilities and the angular distribution of \lilidgs at an energy of 24 MeV was measured and analyzed \cite{shen18}. The spectroscopic amplitudes of these two components were determined to be $0.64\pm0.09$ and $0.74\pm0.09$. The uncertainties of the ANC from the $^{11}$B 3$S_0$ and 2$D_2$ spectroscopic amplitudes were approximately 14\% and 6\%. The details will be published in a further paper. 

\begin{table}[htbp] %add [H] placement to break table across pages
\caption{\label{tab:uncertainties} List of parameters and uncertainty budget for the calculation of the spectroscopic factor ($S_\alpha$) and ANC. The main parameters used in the FRDWBA calculation are shown in the first column. The last column, $\delta_{C^2}$, represents the uncertainty of the ANC from each parameter. Nr and Ni are the normalization factors of the real and imaginary part of the single-folding potential. The subscripts "en" and "ex" represent the entrance and exit channel, respectively. $S_x$ represents the spectroscopic amplitude of $\alpha$ cluster in $x$. It is mentioned that there is only one $\delta_{C^2}$ from each set of $r_0$ and $a$ since $a$ of the bound state is adjusted to reproduce the rms radius of the $\alpha$-cluster wave function. "Angle range" represents the different range of angles used in the fit.}
\begin{ruledtabular}
\begin{tabular}{lccc}
Parameter                             &    Value     &     $\sigma$    &     $\delta_{C^2}$       \\
\hline
$\mathrm{Nr_{en}}$                    &    1.071     &     0.034       &     1.1\%                \\
$\mathrm{Ni_{en}}$                    &    1.388     &     0.049       &     0.5\%                \\
$\mathrm{Nr_{ex}}$                    &    0.744     &     0.063       &     2.9\%                \\
$\mathrm{Ni_{ex}}$                    &    1.56      &     0.10       &     4.9\%                \\
$S_{^{11}\mathrm{B},\mathrm{3S_0}}$   &    0.64      &      0.09       &     13.9\%               \\
$S_{^{11}\mathrm{B},\mathrm{2D_2}}$   &    0.74      &      0.09       &     5.9\%                \\
$r_0$ of $^{16}$O                     &    0.96 fm   &      0.12 fm    &  \multirow{2}{*}{11.4\%} \\
$a$ of $^{16}$O                       &    0.90 fm   &                 &                          \\
$r_0$ of $^{11}$B                     &    0.98 fm   &      0.12 fm    &  \multirow{2}{*}{2.9\%}  \\
$a$ of $^{11}$B                       &    0.60 fm   &                 &                          \\
Statistics                            &              &                 &    10.0\%                 \\
Target thickness                      &              &                 &    7.9\%                 \\
Angle range                           &              &                 &    2.0\%                 \\
Channel radius                        &              &                 &    5.4\%                 \\
\multicolumn{3}{l}{Difference between FRDWBA and CRC}                  &    9.1\%                 \\
\multicolumn{3}{l}{Total uncertainty in ANC and $S_\alpha$}            &    26.1\%                 \\
\end{tabular}
\end{ruledtabular}
\end{table}

The compound nuclear calculations are performed using the Hauser–Fesbach (HF) code CINDY \cite{she73}. The calculations require the optical potentials for the incident and exit channels. These are kept the same as in the FRDWBA calculations described above. The competing channels considered are n, p, $\alpha$ and d populating the corresponding residual nuclei $^{22}$Na, $^{22}$Ne, $^{19}$F and $^{21}$Ne, respectively, in their discrete and continuum states. The number of discrete levels considered is 18 populated from the emission of neutron, proton and alpha, respectively, and 12 from the emission of deuterons. The discrete levels were considered up to the maximum energy available for each channel. The missing energy is considered a continuum and the calculations are performed with a level density parameter of a = A/7 (A is the mass number of the residual nucleus). The spin cutoff parameter was considered as $\sigma=3$ for all nuclei as suggested by Gilbert and Cameron \cite{gil65}. However, the calculation is not very sensitive to the value of $\sigma$. The optical potentials for n+$^{22}$Na, p+$^{22}$Ne, $\alpha$+$^{19}$F and d+$^{21}$Ne are adopted from Wilmore et al. (1964) \cite{wil64}, Perey (1963) \cite{per63} and Daehnick (1980) \cite{dae80}, respectively. The HF calculation is shown by the black dotted line in Fig. 3.

\begin{figure}[htbp]
\centering
\includegraphics[width=\columnwidth]{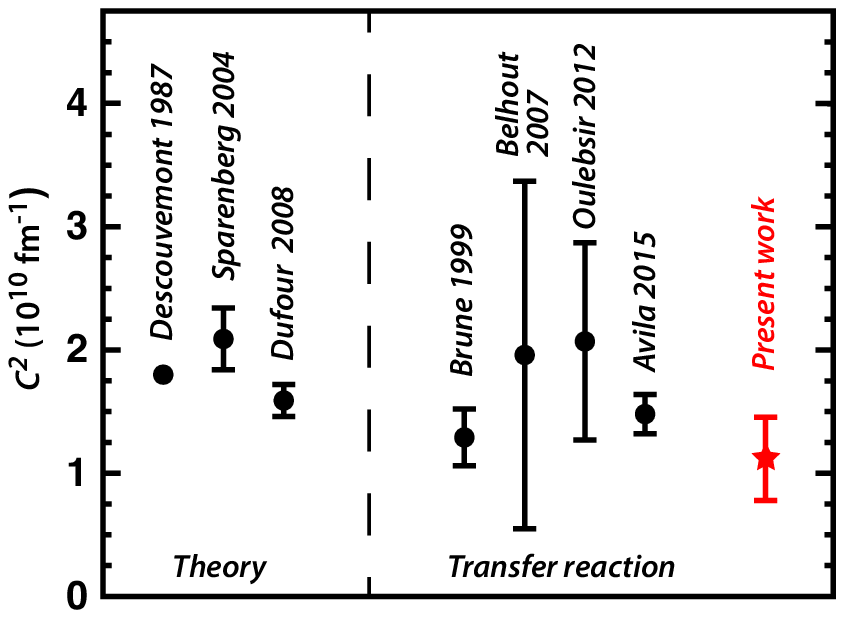}
\caption{(Color online) Comparison of $\alpha$ particle ANCs of the $2^+$ (6.92 MeV) subthreshold state of present and previous works. Theoretical values come from Descouvemont (1987) \cite{des87}, Sparenberg (2004) \cite{spa04} and Dufour et al. (2008) \cite{duf08}. Experimental data are taken from Brune et al. (1999) \cite{bru99} with \Clid~and \Clit~at 2.7-7.0 MeV and 4.75-7.0 MeV, Belhout et al. (2007) \cite{bel07} with \Clid~at 34 and 50 MeV, Oulebsir et al. (2012) \cite{oul12} with \Clit~at 28 and 34 MeV and Avila et al. (2015) \cite{avi15} with \LiCd~at 5, 7, and 9 MeV.}
\label{fig:anc_compare}       % Give a unique label
\end{figure}

It is expected that the FRDWBA model will work best at the most forward angles where there is little compound nucleus reaction contamination. Thus, we fitted the first seven angles where the FRDWBA model best reproduced the experimental data. The spectroscopic factor of the $^{16}$O 6.917 2$^+$ state only changed by 2.0\% with different range of angles used in the fit and this uncertainty was included in the analysis. Figure \ref{fig3} shows the FRDWBA angular distribution of the $^{12}$C($^{11}$B,\,$^{7}$Li)$^{16}$O reaction together with the experimental data. One sees that the FRDWBA calculation reasonably reproduces the experimental data. The spectroscopic factor ($S_\alpha$) was found to be $0.139 \pm 0.034$ by the normalization of the FRDWBA calculation to the experimental angular distribution. The ANC ($C^2$) is related to the spectroscopic factor of the state and the single particle ANC ($b^2$) by the relation ($C^2 = S_\alpha b^2$). The ANC of the 6.917 2$^+$ state in $^{16}$O was extracted to be (1.05 $\pm$ 0.26) $\times 10^{10}$ fm$^{-1}$ using the FRDWBA calculation. The coupled-channel reactions (CRC) calculation was also performed and gave a spectroscopic factor of $0.152 \pm 0.037$ and an ANC of (1.15 $\pm$ 0.28) $\times 10^{10}$ fm$^{-1}$. Finally, the spectroscopic factor and the ANC were determined to be $0.146 \pm 0.038$ and (1.10 $\pm$ 0.29) $\times 10^{10}$ fm$^{-1}$. The difference between the results of the FRDWBA and CRC calculations was treated as a part of the total uncertainty. Table \ref{tab:uncertainties} shows the summary of the parameters and the uncertainty budget. We present the comparison of our ANC and previous works in Fig. \ref{fig:anc_compare}.

% section extraction_of_anc (end)

\begin{table*}[htbp] %add [H] placement to break table across pages
\caption{\label{tab:r-matrix}The resonance parameters used in the $R$-matrix fit of the elastic-scattering of $^{12}$C+$\alpha$ \cite{pla87,tis09} and astrophysical $S_{E2}$ factors. The parameters in the brackets are the fixed resonance parameters taken from Tilley et al. (1993) \cite{til93} except the $\gamma_\alpha^2$ of 6.917 MeV which adopted the value of the present work.}
\begin{ruledtabular}
\begin{threeparttable}
\begin{tabular}{lcccc}
$J_\pi$ & $E_x$ (MeV) & $E_r$ (MeV) & $\gamma_\alpha^2$ or $\Gamma_\alpha$ (keV) & $\Gamma_\gamma$ or $\gamma_\gamma^2$ (keV)\\
\hline
$2^+_1$ & 6.917       & [-0.2449] & $\gamma_\alpha^2=[18.0\pm4.7]$\tnote{a} & $\Gamma_\gamma=[(9.7\pm0.3)\times10^{-5}]$ \\
$2^+_2$ & 9.844       & [2.684]   & $\Gamma_\alpha=0.71\pm19$      & $\Gamma_\gamma=(5.7\pm0.6)\times10^{-6}$ \\
$2^+_3$ & 11.520      & 4.314     & $\Gamma_\alpha=74\pm1$         & $\Gamma_\gamma=(6.1\pm0.2)\times10^{-4}$ \\
$2^+_4$ & 13.020      & 5.833     & $\Gamma_\alpha=112\pm5$        & $\Gamma_\gamma=(7.0\pm2.0)\times10^{-4}$ \\
$2^+_5$ & 15.90       & [8.300]   & $\gamma_\alpha^2=[49.3\pm2.0]$ & $\gamma_\gamma^2=[(1.8\pm0.1)\times10^{-6}]$ \\
$2^+_6$ & 16.443      & [9.281]   & $\gamma_\alpha^2=[1.1\pm0.2]$  & $\gamma_\gamma^2=[(1.9\pm0.1)\times10^{-6}]$ \\
$2^+_7$ & Background  & 22.618    & $\gamma_\alpha^2=4846\pm45$    & $\gamma_\gamma^2=(8.2\pm1.0)\times10^{-3}$ \\
\end{tabular}
\begin{tablenotes}{}
    \footnotesize
    \item[a] Reduced $\alpha$-width from present work.
\end{tablenotes}
\end{threeparttable}
\end{ruledtabular}
\end{table*}

\section{Astrophysical \texorpdfstring{$S_{E2}$}{SE2} factor of the \texorpdfstring{\Cag{}}{12C(a,g)16O} reaction} % (fold)
\label{sec:astrophysical_S_factor}

The astrophysical $S_{E2}$ factor of the ground-state transitions was derived using the best fits on the basis of the $R$-matrix method. The $R$-matrix formulae to fit the cross-sections of the scattering data and S$_{E2}$ under discussion were taken from An et al. (2015)~\cite{an15}. We summarize the formulae here for convenience.

The relevant angular distribution formula for the cross-sections of scattering data is given in Lane et al. (1958) \cite{lan58} by the equation,
\begin{equation}\label{EQ_Diff}
\frac{d\sigma_{\alpha,\alpha{'}}}{d\Omega_{\alpha{'}}}=\frac{1}{(2I_{1}+1)(2I_{2}+1)}
\sum_{ss{'}\nu\nu{'}}^{ }|A_{\alpha{'}s{'}\nu{'},\alpha s\nu}(\Omega_{\alpha{'}})|^{2},
\end{equation}
where the $A_{\alpha{'}s{'}\nu{'},\alpha s\nu}$ are the amplitudes of the outgoing waves.
\begin{widetext}
\begin{equation}\label{EQ_A}
  A_{\alpha{'}s{'}\nu{'},\alpha s\nu}(\Omega_{\alpha^{'}})=\frac{\sqrt{\pi}}{k_{\alpha}}
  [-C_{\alpha{'}}(\theta_{\alpha{'}})\delta_{\alpha{'}s{'}\nu{'},\alpha s\nu}
  +i\sum_{JMll{'}m{'}}^{ }\sqrt{2l+1}(sl\nu0|JM)(s{'}l{'}\nu{'}m{'}|JM)
  T^{J}_{\alpha{'}s{'}l{'},\alpha sl}Y_{m{'}}^{(l{'})}(\Omega_{\alpha{'}})].
\end{equation}
\end{widetext}
Several new quantities have been introduced in Eq.~\eqref{EQ_A} to define the angular dependence of the cross-section. The term $-C_{\alpha{'}}(\theta_{\alpha{'}})$ represents the Coulomb amplitudes, while the $Y_{m'}^{(l')}$ are the spherical harmonics functions. Explicitly, the \textit{T}-matrix is defined by the \textit{R}-matrix components~\cite{lan58}, which characterizes the structure information of the $^{16}$O compound nucleus.

The \textit{R}-matrices are defined as
\begin{equation}\label{EQ_Rmatrix}
  {\bf{R}}^J _{\alpha 's'l',\alpha sl} = \mathop \sum \limits_{\lambda \mu }^N \gamma _{\alpha 's'l'}^J\gamma _{\alpha sl}^J{A_{\lambda \mu }}{\delta _{J{J_0}}},
\end{equation}
where $\gamma _{\alpha 's'l'}^J$ and $\gamma _{\alpha sl}^J$ are the reduced-width amplitude of entrance and exit channel, respectively.

The matrix ${{\mathbf A}}_{{\mathbf \lambda }{\mathbf \mu }}$ is defined by its inverse
\begin{equation}\label{EQ_Ainverse}
  {\left[ {{A^{ - 1}}} \right]_{\lambda \mu }} = \left( {E_\lambda - E} \right){\delta _{\lambda \mu }} - {\Delta }_{\lambda \mu }- \frac{i{\Gamma}_{\lambda \mu}}{2},
\end{equation}
where $E_\lambda$ is the position of resonance level, ${\Delta }_{\lambda \mu }$ is the energy shift, ${\Gamma }_{\lambda \mu }$ is the reduced channel width. The energy shift is
\begin{equation}\label{EQ_Delta1}
  {{\Delta }_{\lambda \mu }} =  - \mathop \sum \limits_{\alpha sl}^N \left( {{{S}_{\lambda \mu }} - {{B}_{\lambda \mu } }} \right)\gamma _{\alpha 's'l'}^{}\gamma _{\alpha sl}^{},
\end{equation}
where $S_{\lambda\mu}$ is the shift factor calculated at the channel radius, and $B_{\lambda\mu}$ is the boundary parameter chosen to equal the shift functions at the energy of the subthreshold state.

For the $^{12}$C($\alpha,\gamma$)$^{16}$O reaction, the cross-section is determined by the following Eq.~\ref{EQ_sigma}, which describes ground state capture in the channel spin representation,
\begin{equation}\label{EQ_sigma}
  {\sigma _{\alpha ',a}} = \frac{\pi }{{k_\alpha ^2}}\mathop \sum \limits_{sl's'lJ}
  \frac{{\left( {2J + 1} \right)}}{{\left( {2{I_1} + 1} \right)\left( {2{I_2} + 1} \right)}}
  {\left| {T_{\alpha 's'l',asl}^J} \right|^2},
\end{equation}
where ${I}_{1}$ and ${I}_{2}$ are the spins of incident particle and target, respectively. The theoretical formulae for error propagation~\cite{an15} are adopted to determine the uncertainty of the extrapolated S factor in our \emph{R}-matrix model fitting.

We repeated the fit of the scattering cross-sections of Plaga et al. (1987)~\cite{pla87} and Tischhauser et al. (2009)~\cite{tis09}, with the same $R$-matrix parameters of An et al. (2015)~\cite{an15}. Level parameters of the reduced-width amplitude of the entrance channel from the fit are in excellent agreement with those reported in An et al. (2015)~\cite{an15}.

For the $R$-matrix fits of S$_{E2}$, we used seven levels associated with the $^{16}$O states at 6.917 (2$^+_1$, $\lambda$ = 1), 9.844 (2$^+_2$, $\lambda$ = 2), 11.520 (2$^+_3$, $\lambda$ = 3) 13.020 (2$^+_4$, $\lambda$ = 4), 15.90 (2$^+_5$, $\lambda$ = 5) and 16.443 MeV (2$^+_6$, $\lambda$ = 6), complemented by a background term (2$^+_7$, $\lambda$ = 7). The levels of $\lambda$ = 4-7 were helpful to reduce the uncertainty produced by the distant levels, and also to subsequently improve the fit precision of S$_{E2}$. The properties of the relevant states are given in Table \ref{tab:r-matrix}. The properties of these states were fixed in the $R$-matrix fits according to Tilley et al. (1993) \cite{til93}, except the reduced $\alpha$ width 2$^+_1$(6.917 MeV), which adopted the value of the present work. The observed reduced $\alpha$ width for the 2$^+_1$(6.917 MeV) state in the present work was given by
\begin{eqnarray}
\gamma_\alpha^2 = \frac{\hbar^2R_c}{2\mu} S_\alpha
\phi(R_c)^2=\frac{\hbar^2}{2\mu R_c} C^2 W(R_c)^2, \label{eq2}
\end{eqnarray}
where $S_\alpha$ and $C^2$ represent the spectroscopic factor and ANC, $\phi(R_c)$ and $W(R_c)$ are the single-particle wave function and the Whittaker function, respectively. The observed reduced $\alpha$ width, $\gamma_\alpha^2$, was converted to the formal channel width in Eq. \ref{EQ_Ainverse} during the R-matrix calculation with the following formula:
\begin{equation}
    \Gamma^{obs}_{\lambda c}=\Gamma_{\lambda c}\left(1+\sum_k\gamma_{\lambda k}^2\frac{dS_k}{dE}\right)^{-1}_{E_\lambda},
\end{equation}
which is Eq. 15 in An et al. (2015)~\cite{an15}. In the present work, $\gamma_\alpha^2$ was extracted to be $18.0\pm4.7$ keV at the channel radius of $R_c$=6.5 fm. This large radius was chosen to reach the Coulomb asymptotic behavior of $\phi(R)$ and was also suggested in Oulebsir et al. (2012) \cite{oul12} and Brune et al. (1999) \cite{bru99}. We also investigated the dependence of the $S_{E2}(300)$ factor on the channel radius by changing $R_c$ from 6.0 fm to 7.0 fm. The uncertainty from $R_c$ was determined to be 5.4\% and was included in the total uncertainty.

The summary of the $R$-matrix parameters in the fits are shown in Table \ref{tab:r-matrix}. Table \ref{tbl-rate} provides fit details such as the normalizations and $\chi^2$ per dataset. In the fits, the procedure was performed according to the same $2^+$-level parameters of this $R$-matrix method \cite{an15} with the astrophysical $S$ factors from previous works \cite{red87,oue96,kun01,ass06,mak09,sch11,pla12}. In general, the data are well fitted where all the energy levels are accurately described. 

\begin{figure}[htbp]
\centering
\includegraphics[width=\columnwidth]{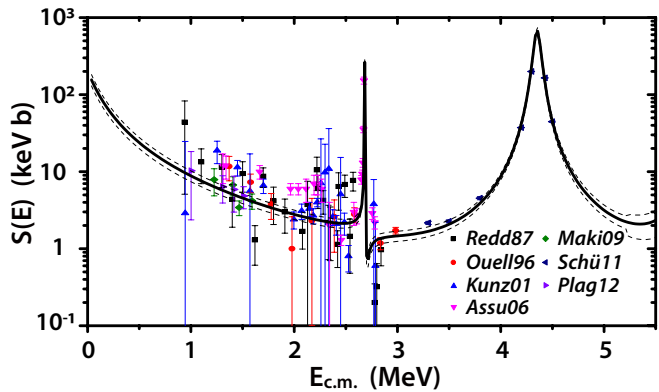}
\caption{(Color online) The comparison of R-Matrix calculations of $E2$ $S$ factor with experimental data \cite{red87,oue96,kun01,ass06,mak09,sch11,pla12}. The solid line is our best $R$-matrix fit using our deduced $\gamma_\alpha^2$ for the 6.917-MeV state, and the dashed lines when using our upper and lower values for $\gamma_\alpha^2$.}
\label{R-Matrix}       % Give a unique label
\end{figure}

\begin{figure}[htbp]
\centering
\includegraphics[width=\columnwidth]{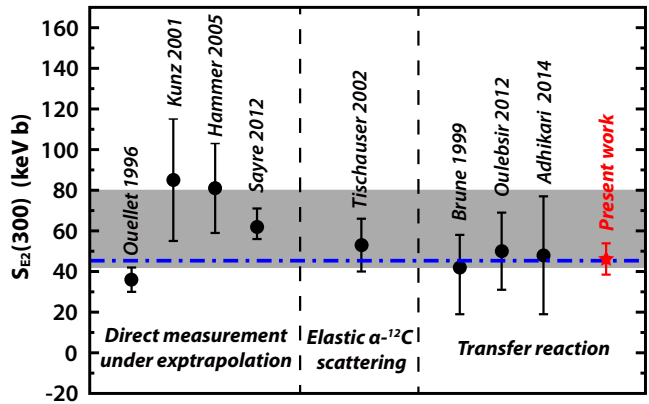}
\caption{(Color online) The $S_{E2}(300)$ comparison of the present work with previous works \cite{tis02,oue96,kun01,ham05,say12,bru99,oul12,adh14}. The grey shadow represents the compilation value of NACRE II (2013) \cite{xu13}. The blue dot-dashed line is the value in deBoer et al. (2017) \cite{deb17}.}
\label{preworks}       % Give a unique label
\end{figure}

The astrophysical S$_{E2}$(300) factor of the ground-state transitions  was derived to be 46.2 $\pm$ 7.7 keV b. The $R$-matrix fits are shown in Fig. \ref{R-Matrix} together with the data from direct measurements \cite{red87,oue96,kun01,ass06,mak09,sch11,pla12}. A comparison of the present S$_{E2}$(300) factor with previous results is shown in Fig. \ref{preworks}. One can see that the present result agrees with the compilation of NACRE II ($61\pm19$ keV b) \cite{xu13} and the most recent compilation by deBoer et al. (2017) (45.3 keV b) \cite{deb17}.

\begin{table}
\caption{\label{tbl-rate} Details of the fit to each dataset of S$_{E2}$, including $\chi^{2}$ contributions from the literature,
normalization parameters and number of data points (ndp) in each $\chi^{2}$ fit. }
\begin{ruledtabular}
\begin{tabular}{lccc}
  Reference & Normalization & $\chi^{2}$ &  $ndp$ \\
\hline
 Plag 2012 \cite{pla12} &   $1.03$ &    $0.660$ &   $  4 $  \\
 Makki 2009 \cite{mak09} &   $1.03$ &    $2.527$ &   $  4 $  \\
 Ouellet 1996 \cite{oue96} &   $0.97$ &    $1.305$ &   $  9 $  \\
 Assun{\c{c}}{\~a}o 2006 \cite{ass06} &   $1.00$ &    $1.968$ &   $ 20 $  \\
 Kunz 2001 \cite{kun01} &   $1.00$ &    $1.034$ &   $ 20 $  \\
 Redder 1987 \cite{red87} &   $1.00$ &    $3.133$ &   $ 24 $    \\
 Sch\"urmann 2011 \cite{sch11} &   $1.03$ &    $5.065$ &   $  7 $  \\
\end{tabular}
\end{ruledtabular}
\end{table}

\section{Conclusion} % (fold)
\label{sec:summary}

In summary, we measured the angular distribution of the \Cbli{} reaction populating the 6.917 MeV 2$^+$ subthreshold state in $^{16}$O using the Q3D magnetic spectrograph at 50 MeV incident energy. The spectroscopic factor and ANC of this state in $^{16}$O were deduced combined a FRDWBA analysis and a CRC analysis, and then used to calculate the reduced $\alpha$ width. The uncertainties in the determined $S_\alpha$, the reduced $\alpha$ width and the ANC were also investigated. Finally, we extracted the astrophysical $S_{E2}(300)$ factor of the ground-state transitions in the \Cag{} reaction to be $46.2\pm7.7$ keV b with the $R$-matrix method. The result for the astrophysical $S_{E2}(300)$ factor confirms the values obtained in various direct and indirect measurements and is in sound agreement with the compilation of NACRE II ($61\pm19$ keV b) \cite{xu13} with the center value lower by 17.5 keV b and in good agreement with the most recent compilation by deBoer et al. (2017) (45.3 keV b) \cite{deb17}, which presents an independent examination of the most important data in nuclear astrophysics.

% section summary (end)

\begin{acknowledgments}
The authors thank the staff of the HI-13 tandem accelerator for the smooth operation of the machine. This work was supported by the National Key Research and Development Project under Grant No. 2016YFA0400502, the National Natural Science Foundation of China under Grants No. 11475264, No. 11490561, No. 11375269 and No. 11327508, and the 973 program of China under Grant No. 2013CB834406. 
\end{acknowledgments}

\bibliography{C12}
\end{document}